\documentclass[aps,preprint,nofootinbib,eqsecnum]{revtex4}%
\pdfoutput=1
\usepackage{amsmath}
\usepackage{amsfonts}
\usepackage{amssymb}
\usepackage{graphicx}
\usepackage{amscd}%
\usepackage{pdfpages}
\providecommand{\U}[1]{\protect\rule{.1in}{.1in}}

\allowdisplaybreaks
\begin{document}
\leftline {USC-10/HEP-B4 \hfill CERN-PH-TH/2010-181}{}{\vskip-1cm}

{\vskip2cm}

\begin{center}
{\LARGE Hunting for TeV Scale Strings at the LHC\footnote{This work was
partially supported by the US Department of Energy under grant number
DE-FG03-84ER40168.}}

{\vskip0.4cm}

\textbf{Itzhak Bars}

{\vskip0.4cm}

\textsl{Department of Physics and Astronomy}

\textsl{University of Southern California,\ Los Angeles, CA 90089-0484 USA}

and

\textsl{Theory Division, Physics Department, CERN, 1211 Geneva 23,
Switzerland}

{\vskip1.0cm} \textbf{Abstract}
\end{center}

In this paper I review the possibility of TeV scale strings that may be
detectable by the Large Hadron Collider (LHC). This possibility was
investigated extensively in a series of phenomenological papers during
1984-1985 in connection with the Superconducting Super Collider (SSC). The
work was mainly based on a model independent systematic parametrization of
scattering amplitudes and cross sections, for Standard Model particles, quarks
and leptons, that were assumed to behave like strings, while gluons, photons,
$W^{\pm},Z$ were taken as elementary. By using Veneziano type beta functions
consistent with crossing symmetry, duality and Regge behavior, bosonic or
fermionic resonances in each channel were included, while the low energy
behavior was matched to effective field theory non-renormalizable interactions
consistent with the Standard Model SU(3)$\times$SU(2)$\times$U(1) gauge
symmetry as well as global flavor and family symmetries. The motivation for
this approach at that time was the possible compositeness of quarks and
leptons but the same phenomenological approach would apply effectively with
the modern additional motivations for TeV scale strings, such as the
hypothesis of D-branes with large extra dimensions. Because some of the main
theoretical and phenomenological work of that time appeared only in the 1984
Snowmass and other proceedings, the results of the investigations have been
inaccessible to most researchers and consequently have been largely forgotten.
Meanwhile similar approaches are being explored by other researchers. Given
the renewed interest in the old results, the purpose of the current paper is
to make them readily available.

\newpage

\section{Strings at the LHC?}

It is everyone's hope that the LHC is on the verge of discovering new physics
at the TeV scale. On the basis of the spectacular success of the Standard
Model, the one sure thing that is expected at the LHC energies is the Higgs
particle or something else that imitates it. Of course clarifying the nature
of the electroweak phase transition will be one of the main tasks at the LHC.
The Standard Model is consistent with several speculative scenarios of new
physics at the TeV scale, including supersymmetry, technicolor, compositeness,
large extra dimensions. While some of these, such as supersymmetry, are better
understood with reliable computational tools, others such as technicolor and
compositeness necessarily remain more obscure due to the presence of an
assumed less understood strong interaction. There are no strong physical
indications that make it necessary for any of these new physical possibilities
to emerge at the TeV scale. Nevertheless none can be eliminated either at this stage.

In this paper I will concentrate on how to hunt for signals of string-like
physics at the TeV scale. The strings I have in mind could be either
elementary, as in the case of large extra dimensions, or flux tubes of a
QCD-like new strong interaction. Spectacular signals would be detected if the
string scale happens to be reachable at the LHC energies. The formalism and
the analysis presented here does not attempt to distinguish between the
underlying physics because it was conceived in 1984 at a time before the first
string revolution, and therefore it concentrates on compositeness. However,
the same approach is easily adapted to the more modern ideas of strings and
branes with large extra dimensions. If string-like signals are detected,
further analysis can in principle be performed to distinguish between specific
phenomena that may be more characteristic of one scenario versus the other
(such as black holes).

For a couple of years before and after the 1984 SSC workshop at Snowmass, I
pursued the phenomenological exploration of string-like physics at the TeV
scale. At that time this study was motivated by the idea that there may be a
new strong interaction at the TeV scale that may hold together preons inside
quarks and leptons. To observe the effects of such a strong interaction I
suggested to look for string-like behavior at the TeV scale, similar to
strings (or flux tubes) that occur in QCD. With this in mind I was the first
to formulate the phenomenological parametrization of cross sections and
amplitudes with Veneziano-type beta functions, to describe all relevant
processes that involved \textquotedblleft composite\textquotedblright\ quarks
and  leptons \cite{snowmass1}\cite{snowmass2}. Then together with
collaborators that included Albright, Barbieri, Dine, Gunion, Hinchliffe,
Eilam and many other phenomenologists and some experimentalists, this type of
phenomenology was pursued for many potential experiments at the SSC. This work
was presented in \cite{snowmass1}-\cite{IBmaryland} most of which was
published in proceedings of conferences that are not easily accessible.
Therefore the papers that appeared in proceedings are being made available as
attachments to the current preprint.

Now there are more motivations, in addition to compositeness, to pursue
string-like physics at the LHC. The idea of D-branes and large extra
dimensions discussed by Arkani-Hamed, Dimopoulos, Dvali, Antoniadis, Randall
and Sundrum \cite{large1}-\cite{large4} has revived the interest for possible
strings at the LHC energies. Although the underlying physics is quite
different than compositeness, the method of phenomenological investigation,
and many of the signals are very similar. If string-like behavior is
miraculously found at the LHC, when enough data becomes available one can
develop tests to distinguish the different physical phenomena as described in
the details of the amplitudes, but the overall phenomenological approach is
common and model independent. Motivated by this view, Cullen, Perelstein and
Peskin \cite{peskin} suggested a phenomenological analysis which was quite
similar to that of \cite{snowmass1}\cite{snowmass2} to investigate the
possibility of strings. More recently L\"{u}st, Stieberger and Taylor
\cite{lust}, discussed an approach which again is similar to the 1984 proposal
in \cite{snowmass1}\cite{snowmass2}, although they proposed more detailed
computations of amplitudes using string theory adapted to the LHC.  Given the
revived interest, and the approaching possibility of new discoveries at the
LHC, the material attached to this preprint may be helpful in the upcoming investigations.

The following are the abstracts of the papers listed in \cite{snowmass1}%
-\cite{IBmaryland}. The interested reader is invited to consult the original
literature. In reading these papers, instead of the SSC now one should
substitute the LHC., and for the most part one can easily substitute the idea
of compositeness by a more general idea that relates to strings at the TeV
scale. Differences in the underlying physics would appear in the details of
the amplitudes.

\begin{itemize}
\item \textit{Physics Above the preon scale, by I. Bars }\cite{snowmass1}: The
preon scale may be comparable or lower than the parton-parton center of mass
energies that will be reached at the SSC. Then we expect resonance, Regge
behavior, diffractive scattering and scaling phenomena to occur in analogy to
the physics of low energy hadronic reactions. Such physical phenomena in the
hadronic world could approximately be described by duality or string-like
behavior (color flux tube confinement). To discuss this type of physics
quantitatively for composite quarks and leptons we propose a model of
scattering amplitudes and cross sections based on Veneziano type beta
functions that correspond to duality diagrams for preons. All dimensionless
parameters are fixed by analogies to low energy hadronic physics. Topics
discussed include the preon scale, the Regge spectrum of composite states,
model independent invariant amplitudes, preon models, preon diagrams and
corresponding duality amplitudes, differential cross sections.

\item \textit{Probing for Preon Structure via Gluons, by C. Albright and I.
Bars }\cite{snowmass2}: Gluons form an important fraction of the partons at
small $x$ in $\bar{p}p$ scattering at SSC energies. Therefore, gluon reactions
at the SSC may be expected to yield important signals for compositeness, if
the preon scale is a few TeV. Here we develop a quantitative method for
estimating many gluon scattering processes. Some of our estimates are model
independent. We also propose explicit formulas for various scattering
amplitudes based on a Veneziano-type beta function model that exhibits
resonances and Regge behavior. Many interesting spectacular signatures are
suggested in the resonance region, where massive vector bosons and/or excited
and exotic quarks and leptons could be produced.

\item \textit{Can the Preon Scale be Low? by I. Bars }\cite{snowmass3}: The
preon scale $\Lambda_{p}$ is bound from below by rare or unobserved processes
\cite{bounds1} and from above by the cosmological abundance of stable heavy
composites \cite{bounds2}. On the other hand composite models can be tested by
the SSC or by low energy experiments only if $\Lambda_{p}$ is allowed to be at
most 5-10 TeV. In search of such models we re-examine some conditions that
must be fulfilled if $\Lambda_{p}$ is small, and point out the possibility of
certain mechanisms that could avoid the dangerous rare processes. In addition,
certain properties of exotic composite particles, their possible role in
breaking the electroweak symmetry and in producing observable signals beyond
the standard model are also discussed.

\item \textit{High p}$_{T}$\textit{ photon production and compositeness at the
SSC, by J.F. Owens, T. Ferbel, M. Dine and I. Bars }\cite{snowmass4}: The
yield for direct photons for p$_{T}\geq1$ TeV is large enough to probe
predictions of conventional QCD as well as to examine consequences of the
compositeness of quarks at the scale of \symbol{126} 5 TeV.

\item  \textit{Low Energy Signals of Composite Models, by R. Barbieri, I.
Bars, M. Bowick, S. Dawson, K. Ellis, H. Haber, B. Holdom, J. Rosner, M.
Suzuki} \cite{snowmass5}: Some signals of compositeness that represent
deviations from the standard model at low energies are discussed. Emphasis is
given to exotic composites, strong P,C violation beyond weak interactions and
small deviations in relations among the parameters of the standard model. Such
effects may be detected at energies obtainable at CERN, LEP and the SSC.

\item  \textit{Searching for quark and lepton compositeness at the SSC, by C.
Albright, I. Bars, K Braun, M. Dine, T. Ferbel, H.J. Lubatti, W.R. Molzon, J.
F. Owens, S.J. Parke, T. R. Taylor, M. P.Schmidt, H. Snow} \cite{snowmass6}:
We examine a variety of issues connected with searching for compositeness at
the SSC. These include effects of resolution, alternative methods of looking
for deviations from QCD predictions, advances of polarized beams, and effects
of compositeness on photon detection. We also consider how physics may look if
the compsiteness scale is as low as a few TeV.

\item  \textit{The effects of quark compositeness at the SSC, by I. Bars and
I. Hinchliffe} \cite{IBhincliff}: The effects of composite quarks on jet cross
sections at the Superconducting Super Collider (SSC) is discussed with
particular emphasis upon the rates for jet energies above the composite scale.
Our main conclusion is that compositeness physics dominates other exotic
physics if the compositeness scale is in the SSC range. Different composite
models are compared in order to examine whether discrimination between them is possible.

\item  \textit{Leptonic signals for compositeness at hadron colliders, by I.
Bars J. F. Gunion and M. Kwan} \cite{IBgunion1}: We consider composite models
in which quarks and leptons have constituents in common. Detailed amplitudes
for subprocesses of the type quark+antiquark $\rightarrow$ lepton+antilepton
are constructed using the presumed similarity between QCD and
compositeness/pre-color interactions. We demonstrate that sensitivity to
compositeness scales, $\Lambda$, as high as 100 TeV to 300 TeV may be
achievable at a supercollider with center-of-mass energy, $\sqrt{s}=40$ TeV.
For moderate values of $\Lambda$ ($\leq$ 15-20 TeV) cross sections are several
orders of magnitude larger than the background Drell-Yan estimate, reflecting
the presence of the new underlying strong interaction. Furthermore, some
models exhibit a resonant structure in lepton anti-lepton pair mass spectra,
near $M_{l\bar{l}}\approx\Lambda$ , corresponding to heavy preon-antipreon
composite states. If $\Lambda$ is in the SSC range, compositeness would
probably dominate the cross sections of most processes and could readily be
explored experimentally. Finally, at $S\bar{p}pS$ energies, $\sqrt{s}%
\approx540$ TeV, observation of lepton-antilepton pair mass spectra with no
deviation from the standard-model Drell-Yan prediction at $M_{l\bar{l}}%
\approx150$ TeV would place limits on $\Lambda$ in the range 1.5 to 3 TeV.

\item \textit{Signals for compositeness in }$e^{+}e^{-}$\textit{ to }%
$e^{+}e^{-}$\textit{ and }$\mu^{+}\mu^{-},$\textit{  by I. Bars J. F. Gunion
and M. Kwan }\cite{IBgunion2}: Theories in which leptons are composite lead to
additional contributions (beyond those from the standard model) to the
amplitudes for e$^{-}$e$^{+}\rightarrow$e$^{-}$e$^{+}$ and e$^{-}$%
e$^{+}\rightarrow$%
$\mu$%
$^{-}$%
$\mu$%
$^{+}$. Detailed models, constructed by analogy between QCD and
compositeness/precolor interactions lead to specific forms for these extra
terms. We demonstrate that compositeness scales M as high as 4--7 TeV may be
probed using e$^{-}$e$^{+}$ collision machines currently available and planned
for the near future. Sensitivity to the type of composite model and its
parity-violation structure is demonstrated. In particular we point out that
there are no standard-model contributions to the scattering e$^{-}$%
e$^{+}\rightarrow$%
$\mu$%
$^{-}$%
$\mu$%
$^{+}$ when the incoming e$^{-}$ and e$^{+}$ both have the same helicity.
Observation of a nonzero cross section in such a helicity scattering state is
prima facie evidence of flavor-changing vector currents in the t channel or
scalar currents connecting the e and
$\mu$
lepton sectors in the s channel.

\item \textit{New physics signatures in polarized }$e^{-}e^{+}$\textit{
experiments, by I. Bars, Gad Eilam, J. Gunion} \cite{barsEilamGunion}: In
$e^{-}e^{+}$ experiments with the $e^{-}$ beam transversally polarized we
compute an asymmetry and show that in the near future it can serve as a probe
of \textquotedblleft new physics\textquotedblright\ up to scales of order 20
TeV. The asymmetry is proportional to the mass of the final state fermion and
hence is largest in the production of the top quark (unless there exists a
heavier fermion). There is  practically no standard model background to this
asymmetry. For example, at center of mass energies of 42 GeV, compositeness
scales of $5,10,15,20$ TeV yield asymmetries of 14.4\%, 3.8\%, 1.7\%, 0.9\%
respectively. We have also found substantial deviations in the unpolarized
cross sections. For example at 200 GeV and $\cos\theta=0.7$ this deviation is
estimated as 355\%, 50\%, 19\%, 10\% at the above compositeness scales. Our
general analysis is applicable to other \textquotedblleft new
physics\textquotedblright\ as well.

\item \textit{Composite quarks and leptons, tests and models, by I. Bars}
\cite{IBmaryland}: The theory and tests of composite quarks and leptons are
reviewed. A model that addresses the puzzle of lepton-quark symmetry within a
family is presented. A dynamical mass generating mechanism is discussed. The
present evidence that requires the compositeness scale to exceed 3 TeV is
reviewed and further tests at the SLC, LEPI and LEPII that would be sensitive
to scales as large as 20-25 TeV are emphasized. Signals that would correspond
to compositeness at SSC energies are described.
\end{itemize}

I hope this compilation of old results would be helpful in the analysis of
data as well as in the construction of new models. One needs to keep an open
mind on the possible underlying physics until the LHC data becomes available.
If and when some candidate events that fit the string-like description are
seen, it would then be very interesting to fit to the general amplitudes in
\cite{snowmass1}\cite{snowmass2} to figure out the details of the underlying physics.

$\allowbreak$

\begin{acknowledgments}
I thank Dieter L\"{u}st for a discussion. I also thank CERN for providing a
stimulating research atmosphere.
\end{acknowledgments}

\newpage
....

\includepdf[pages=-]{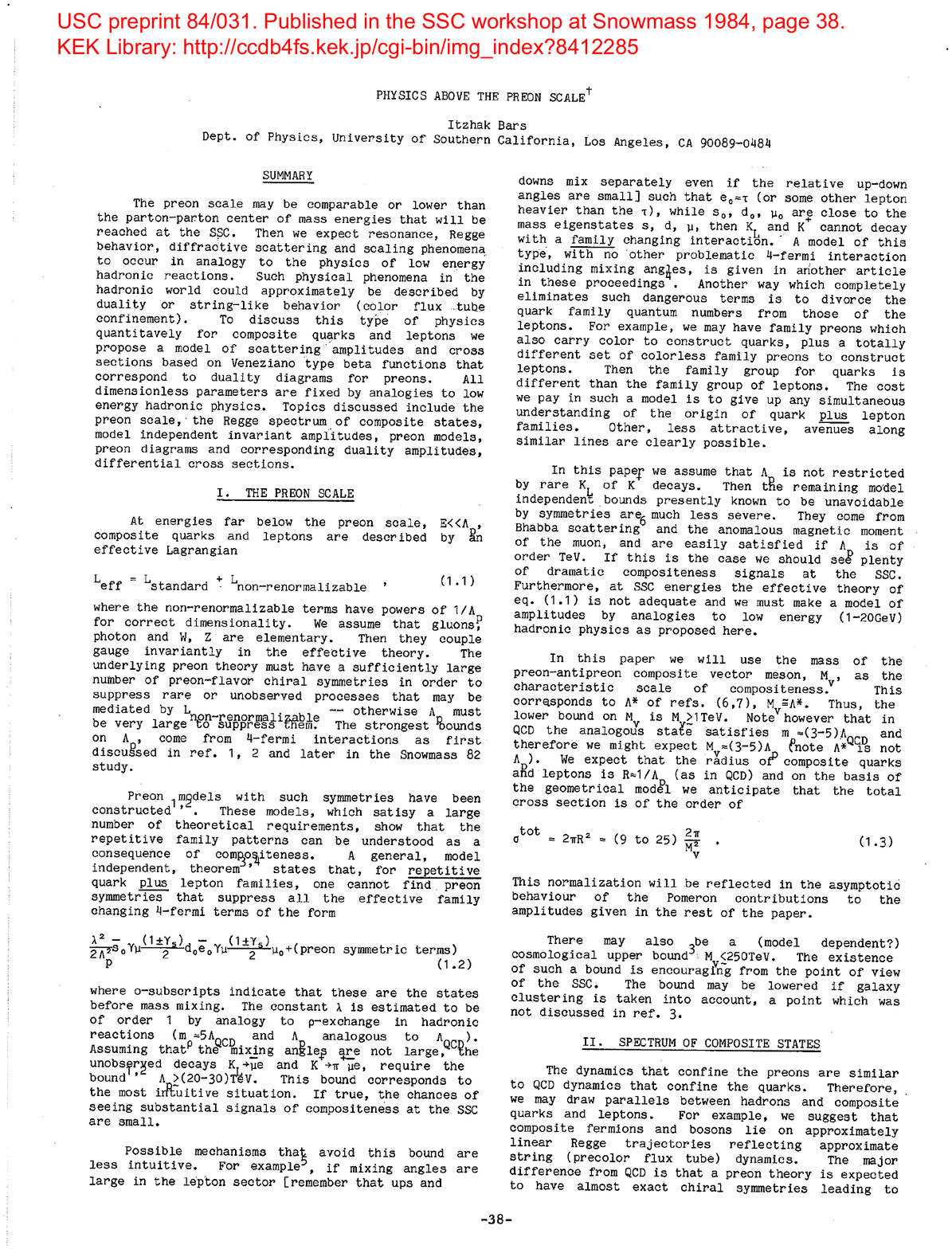} 
\end{document}